\documentstyle[psfig,12pt]{article}
\topmargin - 1.0 true cm
\newcommand{\beq}{\begin{equation}}
\newcommand{\eeq}{\end{equation}}
\def\bea{\begin{eqnarray}}
\def\eea{\end{eqnarray}}
\def\nn{\nonumber}
\def\sss{\scriptscriptstyle}
\def\lft{{\sss L}}
\def\rht{{\sss R}}
\def\roughly#1{\mathrel{\raise.3ex\hbox
{$#1$\kern-.75em\lower1ex\hbox{$\sim$}}}}
\def\lsim{\roughly<}
\def\gsim{\roughly>}
\def\MY{M_{\sss Y}}
\def\MZP{M_{\sss Z'}}
\def\GamZP{\Gamma_{\sss Z'}}

\def\mpla#1#2#3{{\it Mod.\ Phys.\ Lett.} {\bf A#1}, #3 (19#2)}
\def\npb#1#2#3{{\it Nucl.\ Phys.} {\bf B#1}, #3 (19#2)}
\def\plb#1#2#3{{\it Phys.\ Lett.} {\bf B#1}, #3 (19#2)}
\def\prd#1#2#3{{\it Phys.\ Rev.} {\bf D#1}, #3 (19#2)}

\def\prl#1#2#3{{\it Phys.\ Rev.\ Lett.} {\bf #1}, #3 (19#2)}

\begin{document}
\baselineskip=6truemm
\begin{flushright}
LAVAL-PHY-98-12 \\
UdeM-GPP-TH-98-53 \\
McGill-98/26 \\
\end{flushright}

\begin{center}
\bigskip

{\Large \bf Bilepton Production at Hadron Colliders}
\bigskip

B. Dion$^{a}$,
T. Gr\'egoire$^{b,}$\footnote{Address after Sept.\ 1, 1998: Department of
Physics, University of California, Berkeley, California 94720, USA},
D. London$^{b,}$\footnote{london@lps.umontreal.ca},
L. Marleau$^{a,}$\footnote{lmarleau@phy.ulaval.ca},
and H. Nadeau$^{c,}$\footnote{nadeau@hep.physics.mcgill.ca}

\end{center}


\begin{flushleft}
~~~~~~~~~~~~~$a$: {\it D\'epartement de Physique, Universit\'e Laval,
Qu\'ebec QC Canada, G1K 7P4} \\
~~~~~~~~~~~~~$b$: {\it Laboratoire Ren\'e J.-A. L\'evesque,
Universit\'e de Montr\'eal,}\\
~~~~~~~~~~~~~~~~~{\it C.P. 6128, succ. centre-ville, Montr\'eal, QC,
Canada, H3C 3J7}
\\
~~~~~~~~~~~~~$c$: {\it Physics Department, McGill University,}\\
~~~~~~~~~~~~~~~~~{\it 3600 University St., Montr\'eal, QC, Canada, H3A 2T8}
\end{flushleft}

\begin{center}

\bigskip
(\today)
\bigskip\\

{\bf Abstract}

\end{center}

\begin{quote}

We examine, as model-independently as possible, the production of bileptons
at hadron colliders. When a particular model is necessary or useful, we
choose the 3-3-1 model. We consider a variety of processes: $q{\bar q} \to
Y^{++} Y^{--}$, $u{\bar d} \to Y^{++} Y^{-}$, ${\bar u}d \to Y^+Y^{--}$,
$q{\bar q} \to Y^{++} e^-e^-$, $q{\bar q} \to \phi^{++} \phi^{--}$, $u{\bar
d} \to \phi^{++} \phi^{-}$, and ${\bar u} d \to \phi^{+}\phi^{--}$, where
$Y$ and $\phi$ are vector and scalar bileptons, respectively. Given the
present low-energy constraints, we find that at the Tevatron, vector
bileptons are unobservable, while light scalar bileptons ($M_\phi \lsim
300$ GeV) are just barely observable. At the LHC, the reach is extended
considerably: vector bileptons of mass $\MY \lsim 1$ TeV are observable, as
are scalar bileptons of mass $M_\phi \lsim 850$ GeV.

\end{quote}

\newpage


\section{Introduction}

All models of physics beyond the standard model (SM) predict the existence
of new particles. One of the more exotic of these is the bilepton
\cite{Cuypers}, a particle of lepton number 2. Bileptons occur in a variety
of models of new physics. For example, the gauge bosons of $SU(15)$ grand
unified theories \cite{SU15} include vector bileptons, as do models with an
$SU(3)_c \times SU(3)_\lft \times U(1)$ gauge symmetry (known as the 3-3-1
model) \cite{331}. Scalar bileptons can be found in models with an extended
Higgs sector such as left-right symmetric models, or models in which
Majorana neutrino masses are generated.

Since bileptons couple to a pair of leptons, there are significant
constraints on their masses and couplings from low-energy data
\cite{Cuypers}. The most stringent constraints on doubly-charged scalar and
vector bileptons come from searches for muonium-antimuonium conversion:
$M_\lft > 1.7$--$3.3 \, \lambda$ TeV, where $M_\lft$ is the mass of the
bilepton, and $\lambda$ its coupling to leptons. The constraints on
singly-charged bileptons are due to experimental limits on $\mu_\rht \to
e\nu\nu$ and $\nu_\mu \to \nu_e$ oscillations, and are slightly weaker:
$M_\lft > 1$--$2 \, \lambda$ TeV. Thus, for couplings $\lambda \sim 1$,
bileptons are generically constrained to have a mass greater than $\sim 1$
TeV.

There has been considerable work examining the prospects for the detection
of bileptons at future colliders. Because bileptons couple principally to
leptons, it is only natural that this work has concentrated mainly on
colliders involving at least one lepton beam \cite{colliders}. However,
bileptons also couple to the photon and $Z$, and so could be produced at
hadron colliders. Curiously, the possibility of detecting bileptons at
future hadron colliders has been little studied in the literature. For a
high-energy $pp$ collider, the production of the vector bileptons of the
3-3-1 model and the scalar bileptons in left-right symmetric models has
been calculated in Refs.~\cite{bilepSSC} and \cite{bilepLRS}, respectively.
But nobody has attempted to perform a systematic study of bilepton
production at hadron colliders. This is the purpose of this paper.

It is clear from the start that $e^+e^-$ and $e^-e^-$ colliders potentially
have a great advantage over hadron colliders for detecting bileptons.
High-energy Bhabha and M{\o}ller scattering receive huge corrections from
virtual bilepton exchange. If no deviation from the SM is seen, this will
constrain the mass of the bilepton to be $M_\lft \gsim 50 \sqrt{s} \,
\lambda$ TeV \cite{Cuypers}. In other words, depending on the value of the
coupling $\lambda$, the reach of $e^+e^-$ and $e^-e^-$ colliders for
bilepton detection potentially extends far beyond their centre-of-mass
energies. On the other hand, it is equally evident that this reach depends
crucially on the value of the $\lambda$. The advantage of hadron colliders,
in which bilepton production is due mainly to the $s$-channel exchange of
neutral gauge bosons, is that the cross section depends only on gauge
couplings, so that the reach is independent of the coupling $\lambda$. (The
same holds true for direct searches in $e^+e^-$ colliders.) Thus, even
though $e^+e^-$ and $e^-e^-$ colliders are potentially better tools for
bilepton detection, it is still worthwhile to consider hadron colliders.

Ideally, the study of bilepton production at hadron colliders should be
completely model-independent. Unfortunately, this is not possible. Consider
the process $q{\bar q} \to Y^{++} Y^{--}$, where $Y^{++}$ is a
doubly-charged vector bilepton. If one calculates the cross section for
this process using only $s$-channel $\gamma$ and $Z$ exchange, one finds
that the cross section grows with the centre-of-mass energy, i.e.\
unitarity is violated. This is not surprising. The vector bileptons are the
gauge bosons of a larger gauge group, which necessarily contains at least
one new neutral $Z'$ boson. It is only through the inclusion of the
$s$-channel $Z'$ exchange that unitarity is restored\footnote{This is
completely analogous to what happens in the SM. Using only photon exchange
the cross section for $e^+e^- \to W^+W^-$ violates unitarity. Unitarity is
restored when the contribution of the neutral gauge boson associated with
the $W$ --- the $Z$ --- is also included in the calculation.}. Thus, for
vector bilepton production at hadron colliders, it is necessary to choose a
model in which to perform the calculation. In this paper we choose the
3-3-1 model \cite{331}. Strictly speaking, our results apply only to this
model. However, we expect that the order of magnitude of the cross sections
will hold in any model containing vector bileptons.

For scalar bileptons, one does not have the same problems with unitarity
violation. Thus, the calculations of the cross sections for scalar bilepton
production at hadron colliders can be performed without a knowledge of the
underlying theory. That is, one can consider only $s$-channel $\gamma$ and
$Z$ exchange, which is indeed what we do. However, it must be remembered
that any particular model may contain $Z'$ bosons, which can affect the
cross sections (especially if the $Z'$ can be produced on shell). In this
sense, the cross sections for scalar bilepton production presented in this
paper should be considered as lower bounds -- in a given model, these cross
sections may be enhanced due to the exchange of other particles. (It is
also conceivable that the cross sections could be decreased, due to
cancellations between the $Z'$ and $\gamma/Z$ contributions. However, in
general, this requires fine-tuning.)

We discuss vector bilepton production at both the Tevatron and the Large
Hadron Collider (LHC) in Section 2. We consider the production of two real
bileptons ($q{\bar q} \to Y{\bar Y}$), as well as the case where one of the
bileptons is virtual ($q{\bar q} \to Yee$). In Section 3 we turn to scalar
bilepton production. We conclude in Section 4.


\section{Vector Bileptons}

The basic process describing the production of vector bileptons at hadron
colliders is $q{\bar q} \to Y^{++} Y^{--}$, where $Y^{++}$ is a
doubly-charged vector bilepton. Ideally we would like to study this process
model-independently. So, as a first step, we compute the cross section
based on the $s$-channel exchange of the $\gamma$ and $Z$ only. However,
the calculation reveals that this cross section grows as $s$, where
$\sqrt{s}$ is the centre-of-mass energy. This signals a violation of
unitarity, indicating that there are other important contributions to this
process which have not been taken into account.

But it is clear what is happening here. Vector bileptons are the gauge
bosons of some larger gauge group, which must necessarily include new
neutral $Z'$ gauge bosons. These $Z'$ bosons will also contribute to
$q{\bar q} \to Y^{++} Y^{--}$, and their inclusion must restore unitarity.
Thus, it is not possible to perform a model-independent study of vector
bilepton production at hadron colliders. It is necessary to choose a
particular model, so as to be able to include the $Z'$ (and possibly other)
contributions which restore unitarity.

For computational purposes, we therefore choose the simplest extension of
the SM which contains bileptons, namely the model in which the $SU(2)_\lft$
gauge group is expanded to $SU(3)_\lft$, giving an $SU(3)_c \times
SU(3)_\lft \times U(1)$ gauge symmetry (the 3-3-1 model) \cite{331}. Within
the minimal 3-3-1 model, the calculation of the cross section for double
vector-bilepton production at hadron colliders has been performed in
Ref.~\cite{bilepSSC}. In order to be as general as possible, we consider
bilepton production in the nonminimal 3-3-1 model. We also examine the
possibility of single production of vector bileptons. Note that, although
these calculations are clearly not model-independent, we do expect that the
order of magnitude of the cross sections will be the same in any model
containing vector bileptons. After all, any model with vector bileptons
will have the same types of contributions to $q{\bar q} \to Y^{++} Y^{--}$
as one finds in the the 3-3-1 model.

We begin this section with a review of the 3-3-1 model.

\subsection{The 3-3-1 model}

We present here the main features of the 3-3-1 model, concentrating
principally on those ingredients which are necessary for our calculation.
For more details, we refer the reader to Ref.~\cite{DNg}.

In the 3-3-1 model, the gauge group is $SU(3)_c \times SU(3)_\lft \times
U(1)_{\sss X}$, in which the coupling constants of $SU(3)_\lft$ and
$U(1)_{\sss X}$ are denoted $g$ and $g_{\sss X}$, respectively. The group
$SU(3)_\lft \times U(1)_{\sss X}$ is broken to $SU(2)_\lft \times
U(1)_{\sss Y}$ when an $SU(3)_\lft$-triplet scalar gets a vacuum
expectation value. The matching of the gauge coupling constants at this
breaking scale yields the relation
\beq
{g_{\sss X}^2 \over g^2} = {6 \sin^2 \theta_w \over 1 - 4 \sin^2 \theta_w}
~.
\eeq

When $SU(3)_\lft \times U(1)_{\sss X}$ is broken to $SU(2)_\lft \times
U(1)_{\sss Y}$, there are five exotic gauge bosons which acquire masses.
They are the doubly- and singly-charged bileptons $Y^{++}$, $Y^+$ and
their antiparticles, along with a new neutral $Z'$ gauge boson. When the
minimal Higgs structure is used to break the symmetry, there is a relation
between the masses: 
\beq
\label{mincon}
{M_{\sss Y} \over M_{Z'}} = {\sqrt{3 (1 - 4 \sin^2 \theta_w)} \over 2
\cos\theta_w} ~,
\eeq
where $M_{\sss Y^+} \simeq M_{\sss Y^{++}} \equiv M_{\sss Y}$. In this
paper, in order to be as general as possible, we do not assume the minimal
Higgs structure. Hence we allow $M_{\sss Y}$ and $M_{\sss Z'}$ to vary
independently of one another.

The fermions transform under the 3-3-1 symmetry as follows:
\bea
\psi_{1,2,3} = \left( \matrix{ e \cr \nu_e \cr e^c \cr} \right) ~,~
\left( \matrix{ \mu \cr \nu_\mu \cr \mu^c \cr} \right) ~,~
\left( \matrix{ \tau \cr \nu_\tau \cr \tau^c \cr} \right)
& : & (1,3^*,0) ~, \nonumber\\
Q_{1,2} = \left( \matrix{ u \cr d \cr D_1 \cr} \right) ~,~
\left( \matrix{ c \cr s \cr D_2 \cr} \right) 
& : & (3,3,-{1\over 3}) ~, \nonumber\\
Q_3 = \left( \matrix{ t \cr b \cr T \cr} \right) & : &
(3,3^*,{2\over 3}) ~, \nonumber\\
d^c, s^c, b^c & : & (3^*, 1, {1\over 3}) ~,~ \nonumber \\
u^c, c^c, t^c & : & (3^*, 1, -{2\over 3}) ~,~ \nonumber \\
D_1^c, D_2^c & : & (3^*, 1, {4\over 3}) ~,~ \nonumber \\
T^c & : & (3^*, 1, -{5\over 3}) ~.
\eea
In this model there are three new, exotic quarks of charge $-{4\over 3}$
($D_{1,2}$) and ${5\over 3}$ ($T$). Here anomaly cancellation
takes place among all three generations, in contrast to the SM, where the
anomalies are cancelled within each generation.

We write the Feynman rules for the couplings of the neutral gauge bosons
and the fermions as 
\beq
i g \left[ c_{\sss N}^{f_\lft} \, \gamma^\mu \, {1 - \gamma_5 \over 2} + 
c_{\sss N}^{f_\rht} \, \gamma^\mu \, {1 + \gamma_5 \over 2} \right],
\eeq
where $N = \gamma, Z, Z'$. The couplings of the photon and $Z$ to the
fermions are as in the SM, while those of the $Z'$ are
\beq
c_{\sss Z'}^{f_{\lft,\rht}} = {1 \over 2 \sqrt{3} \cos\theta_w \sqrt{1 - 4
\sin^2 \theta_w} } \, d_{\sss Z'}^{f_{\lft,\rht}} ~,
\eeq
where the $d_{\sss Z'}^{f_{\lft,\rht}}$ are given in
Table~\ref{Z'couplings}.

\begin{table}
\hfil
\vbox{\offinterlineskip
\halign{&\vrule#&
 \strut\quad#\hfil\quad\cr
\noalign{\hrule}
height2pt&\omit&&\omit&&\omit&\cr
& \omit && $d_{\sss Z'}^{f_\lft}$ && $d_{\sss Z'}^{f_\rht}$ & \cr
height2pt&\omit&&\omit&&\omit&\cr
\noalign{\hrule}
height2pt&\omit&&\omit&&\omit&\cr
& $u,c$ && $-(1-2\sin^2\theta_w)$ && $+4\sin^2\theta_w$ & \cr
& $d,s$ && $-(1-2\sin^2\theta_w)$ && $-2\sin^2\theta_w$ & \cr
& $\ell^-$ && $+(1-4\sin^2\theta_w)$ && $+2(1-4\sin^2\theta_w)$ & \cr
& $\nu_\ell$ && $+(1-4\sin^2\theta_w)$ && \omit & \cr
& $t$ && $+1$ && $+4\sin^2\theta_w$ & \cr
& $b$ && $+1$ && $-2\sin^2\theta_w$ & \cr
& $D_1,D_2$ && $+2(1-5\sin^2\theta_w)$ && $-8\sin^2\theta_w$ & \cr
& $T$ && $-2(1-6\sin^2\theta_w)$ && $+10\sin^2\theta_w$ & \cr
height2pt&\omit&&\omit&&\omit&\cr
\noalign{\hrule}}}
\caption{Values of the $d_{\sss Z'}^{f_\lft}$ and $d_{\sss Z'}^{f_\rht}$
parameters which define the $Z'$ couplings to fermions.}
\label{Z'couplings}
\end{table}

\begin{figure}[t]
\centerline{
\mbox{\psfig{figure=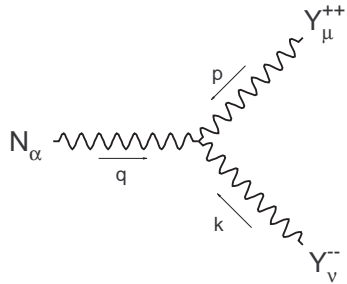,width=2.0in}}
}
\caption{Momentum and Lorentz-index assignments for the
$N$--$Y^{++}$--$Y^{--}$ vertex (Eq.~\protect\ref{NYYvertex}).}
\label{NYYvtx}
\end{figure}

We now turn to the trilinear gauge-boson vertices. The Feynman rule for
the $N$--$Y^{++}$--$Y^{--}$ vertex (see Fig.~\ref{NYYvtx}) is
\beq
\label{NYYvertex}
i g \, c_{\sss N}^{\sss Y} \left( g^{\mu\nu} (k-p)^\alpha + g^{\nu\alpha}
(q-k)^\mu + g^{\alpha\mu} (p-q)^\nu \right) ~,
\eeq
where
\beq
c_\gamma^{\sss Y} = 2 \sin\theta_w ~,~~~~
c_{\sss Z}^{\sss Y} = {1 - 4 \sin^2 \theta_w \over 2 \cos \theta_w} ~,~~~~
c_{\sss Z'}^{\sss Y} = - {\sqrt{3} \over 2} \sqrt{1 - 3 \tan^2 \theta_w} ~.
\eeq

Finally, the data on muonium-antimuonium conversion constrain
doubly-charged vector bileptons to satisfy $\MY > 1.7 \, \lambda$ TeV
\cite{Cuypers}, where $\lambda$ is the bilepton coupling to leptons. In the
3-3-1 model, we have $\lambda = g/\sqrt{2}$, which implies that the lower
limit on the vector bilepton mass is $\MY > 740$ GeV. For singly-charged
vector bileptons, the limits on $\mu_\rht \to e\nu\nu$ yield $\MY > 440$
GeV.

As for the $Z'$, its couplings to quarks are enhanced relative to its
couplings to leptons. Thus, limits on the mass of the $Z'$ come mainly from
low-energy experiments such as neutrino-quark scattering and atomic parity
violation. These constrain $\MZP \gsim 550$ GeV \cite{DNg}.

With this information we can now proceed to the calculation of the cross
sections for the production of bileptons.

\newpage
\subsection{$q{\bar q} \to Y^{++} Y^{--}$}

\begin{figure}[t]
\centerline{
\mbox{\psfig{figure=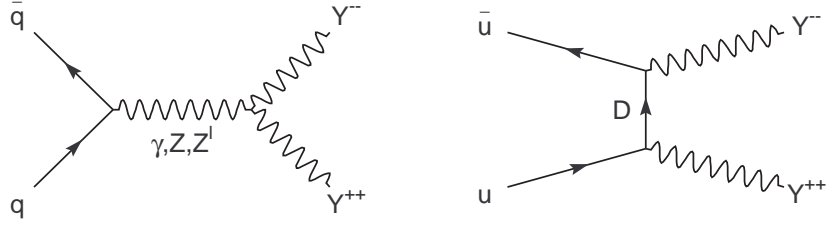,width=4.5in}}
}
\caption{Feynman diagrams contributing to $q{\bar q} \to Y^{++} Y^{--}$ in
the 3-3-1 model.}
\label{Feyndiag}
\end{figure}

The process $d{\bar d} \to Y^{++} Y^{--}$ receives contributions from
$s$-channel $\gamma$, $Z$ and $Z'$ exchange (see Fig.~\ref{Feyndiag}). The
amplitude-squared is
\bea
{1\over 4}\sum_{\rm spins} |{\cal M}|^2 & = & \sum_{\sss N,N'} {1\over 2} \,
{g^4 \, c_{\sss N}^{\sss Y} \, c_{\sss N'}^{\sss Y} \, ( c_{\sss N}^{d_\lft} 
\, c_{\sss N'}^{d_\lft} + c_{\sss N}^{d_\rht} \, c_{\sss N'}^{d_\rht} )
\over ({\hat s} - M_{\sss N}^2 + i \Gamma_{\sss N} M_{\sss N})
({\hat s} - M_{\sss N'}^2 - i \Gamma_{\sss N'} M_{\sss N'}) } \times \nn \\
&~& {\hskip-2truecm}
\left[ \left( - 6 \MY^2 {\hat s} + {{\hat s}^4 \over 8 \MY^4} \right) 
\left(1 - \cos^2\theta \right) + {{\hat s}^3 \over \MY^2} \left(1 +
\cos^2\theta \right) - {9\over 2} {\hat s}^2 \left(1 + {7\over 9}
\cos^2\theta \right) \right], 
\eea
where $N,N' = \gamma,Z,Z'$, and ${\hat s}$ is the centre-of-mass energy of
the $q{\bar q}$ system (not to be confused with $s$, the centre-of-mass
energy of the collider). It is straightforward to verify that this
expression does not violate unitarity. In the limit as $\sqrt{{\hat s}} \to
\infty$, $M_{\sss N}$ and $\Gamma_{\sss N}$ are negligible. But in this
limit the cross section vanishes since $\sum_{\sss N} c_{\sss N}^{\sss Y}
c_{\sss N}^{d_{\lft,\rht}} = 0$. Thus we see that, as expected, the
inclusion of the new contributions (in this case a $Z'$) restores
unitarity.

One quantity which appears in the above expression, and which we have not
yet discussed, is $\GamZP$. The $Z'$ may decay to the exotic quarks $D_i$
and $T$, depending on their mass. As in Ref.~\cite{bilepSSC}, we assume
that $m_{\sss Q} = 600$ GeV ($Q=D_i,T$). Furthermore, the $Z'$ may decay to
the light scalars of the Higgs sector. But since we are considering a
general nonminimal 3-3-1 model, we have not specified the Higgs sector. In
the minimal model, the partial width of the $Z'$ into light scalars is
roughly 10\% of its width into the light SM quarks. For simplicity, here we
assume that even with a nonminimal Higgs sector the partial width into
scalars is the same as in the minimal model. Note that this assumption does
not have strong consequences -- the cross sections do not depend much on
this partial width.

The process $u{\bar u} \to Y^{++} Y^{--}$ is a bit more complicated. In
addition to the $s$-channel $\gamma$, $Z$ and $Z'$ contributions, there is
a diagram in which a $D_1$ quark is exchanged in $t$-channel (see
Fig.~\ref{Feyndiag}). We denote these two amplitude types as ${\cal
M}_1$ and ${\cal M}_2$, respectively. The amplitude-squared is then the sum
of the following three terms:
\newpage
\bea
\label{uuamp1}
{1\over 4}\sum_{\rm spins} |{\cal M}_1|^2 & = & \sum_{\sss N,N'} {1\over 2}
\, {g^4 \, c_{\sss N}^{\sss Y} \, c_{\sss N'}^{\sss Y} \, ( c_{\sss
N}^{u_\lft} \, 
c_{\sss N'}^{u_\lft} + c_{\sss N}^{u_\rht} \, c_{\sss N'}^{u_\rht} ) \over
({\hat s} - M_{\sss N}^2 + i \Gamma_{\sss N} M_{\sss N})
({\hat s} - M_{\sss N'}^2 - i \Gamma_{\sss N'} M_{\sss N'}) } \times \nn \\
&~& {\hskip-2.5truecm}
\left[ \left( - 6 \MY^2 {\hat s} + {{\hat s}^4 \over 8 \MY^4} \right) 
\left(1 - \cos^2\theta \right) + {{\hat s}^3 \over \MY^2} \left(1 +
\cos^2\theta \right) - {9\over 2} {\hat s}^2 \left(1 + {7\over 9}
\cos^2\theta \right) \right], 
\eea
\bea
\label{uuamp2}
{1\over 4}\sum_{\rm spins} |{\cal M}_2|^2 & = & {g^4\over 16} \, {1 \over
({\hat t} - M_{\sss D}^2)^2 } \Biggl\{ -{\hat s} ( 1 - 5 \cos^2 \theta )
\MY^2 \nn \\ 
& ~ & {\hskip-2truecm}
- {\hat s}^2 \left( {7\over 4} + {21\over 4} \cos^2\theta + \cos^4\theta
\right) + {{\hat s}^3 \over 2 \MY^2} ( 1 + \cos^2 \theta )^2 + {{\hat s}^4
\over 16 \MY^4} (1 - \cos^4\theta ) \nn \\
& ~ & {\hskip-2truecm}
\left.
- {\hat s}^2 \beta (3 \cos\theta + \cos^3\theta) + {{\hat s}^3 \over 4
\MY^2} \beta (5 \cos\theta + 3 \cos^3\theta) + {{\hat s}^4 \over 8 \MY^4}
\beta \cos\theta \sin^2\theta \right\} ~,
\eea
and
\bea 
\label{uuamp3}
{1\over 4}\sum_{\rm spins} {\cal M}_1 {\cal M}_2^* + h.c. & = & 
\sum_{\sss N} g^4 { c_{\sss N}^{\sss Y} \, c_{\sss N}^{u_\lft} \, 
({\hat s} - M_{\sss N}^2) \over 
\left[ \left( {\hat s} - M_{\sss N}^2 \right)^2 +  
\left( \Gamma_{\sss N} M_{\sss N} \right)^2 \right] 
({\hat t} - M_{\sss D}^2) } 
\left\{ -{3\over 2} {\hat s} \MY^2 \sin^2\theta 
\right. \nn \\
& ~ & {\hskip-2truecm}
- {{\hat s}^2 \over 8}(9 + 7 \cos^2\theta)
+ {{\hat s}^3 \over 4 \MY^2} (1 + \cos^2\theta) 
- {1\over 4} {\hat s}^2 \beta \cos\theta (3 + \cos^2 \theta) 
\nn \\
& ~ & {\hskip-2truecm}
\left. 
+ {{\hat s}^4 \over 32 \MY^4} \sin^2 \theta
+ {{\hat s}^3 \over 16 \MY^2} \beta (5 \cos\theta + 3
\cos^3\theta) + {{\hat s}^4 \over 32 \MY^4} \beta \cos\theta \sin^2\theta
\right\} ~,
\eea
where $\beta \equiv \sqrt{1 - 4 \MY^2/ {\hat s}}$.

We obtain the cross section for bilepton production at hadron
colliders by convoluting the above expressions with the CTEQ3M
structure functions \cite{CTEQ3M}. We consider both the Tevatron
($\sqrt{s}=1.8$ TeV) and the LHC ($\sqrt{s}=14$ TeV). The present
luminosity at the Tevatron is 100 $pb^{-1}$/year (run 1), and this is
expected to be increased to at least 2 $fb^{-1}$/year in 2002 (run
2)\footnote{Strictly speaking, run 2 involves not only a luminosity
  increase, but also an increase in energy from 1.8 TeV to 2.0 TeV.
  For simplicity, in the figures we continue to take $\sqrt{s} = 1.8$
  TeV at the Tevatron for both runs. However, we have also performed
  the calculations for 2.0 TeV. Although the cross sections are
  increased by a factor of 1.5 to 2, this does not affect our
  conclusions significantly. For the various processes, we indicate in
  the text the effects of using $\sqrt{s} = 2.0$ TeV instead of 1.8
  TeV for run 2.}
The design luminosity at the LHC is 10 $fb^{-1}$/year. Since we have
not included a rapidity cut on the produced particles, nor have we
taken into account detection efficiency, as a figure of merit we
therefore (conservatively) require 25 events for discovery. This
corresponds to a cross section of 0.25 $pb$ (run 1) or 12.5 $fb$ (run
2) at the Tevatron, and 2.5 $fb$ at the LHC.

\begin{figure}[t]
\centerline{
\hskip-1truecm
\mbox{\psfig{figure=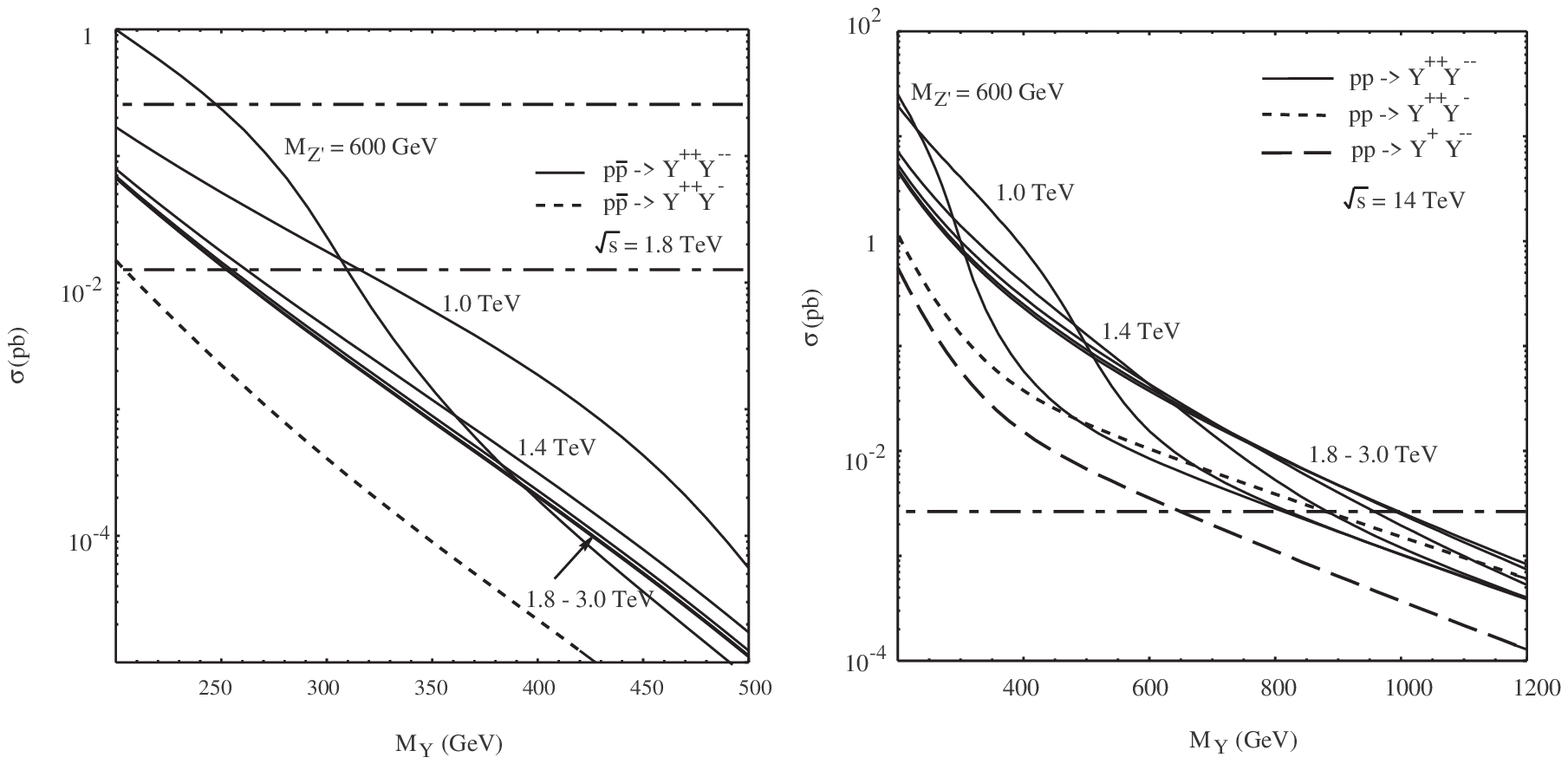,width=7.0in}}
}
\vskip-5.1truein
\caption{Cross sections for $Y^{++}Y^{--}$, $Y^{++}Y^-$ and $Y^{--}Y^+$
production at the Tevatron ($\protect\sqrt{s} = 1.8$ TeV) and at the LHC
($\protect\sqrt{s} = 14$ TeV) as a function of $\MY$, for various values of
$\MZP$. The horizontal lines indicate the cross sections required for
discovery: 0.25 $pb$ (Tevatron, run 1), 12.5 $fb$ (Tevatron, run 2), and
2.5 $fb$ (LHC).} 
\label{YY}
\end{figure}

The results for $Y^{++}Y^{--}$ production are shown in Fig.~\ref{YY},
where the cross sections are plotted as a function of $\MY$ for
various $Z'$ masses. (In Ref.~\cite{bilepSSC}, the production cross
section is calculated within the minimal 3-3-1 model, in which the
condition of Eq.~\ref{mincon} between $\MY$ and $\MZP$ is assumed.
When we impose this condition, we find that we do indeed reproduce the
results of Ref.~\cite{bilepSSC}.) For the Tevatron (run 1), we see
that only bileptons of mass $\MY \lsim 250$ GeV are observable if
$\MZP \ge 600$ GeV. This increases modestly to $\MY \lsim 320$ GeV in
run 2 (for $\sqrt{s} = 2$ TeV, this upper limit becomes 360 GeV).
However, as explained above, in all cases this bilepton mass range has
already been ruled out.  We therefore conclude that, given the present
constraints on $\MY$ from low-energy data, vector bileptons cannot be
observed at experiments at the Tevatron.

At the LHC, on the other hand, bileptons of mass $\MY \lsim 1$ TeV are
observable. Since the vector bilepton mass is presently constrained to be
$\MY > 740$ GeV, this means that there is a window of observability.

\begin{figure}[t]
\centerline{
\mbox{\psfig{figure=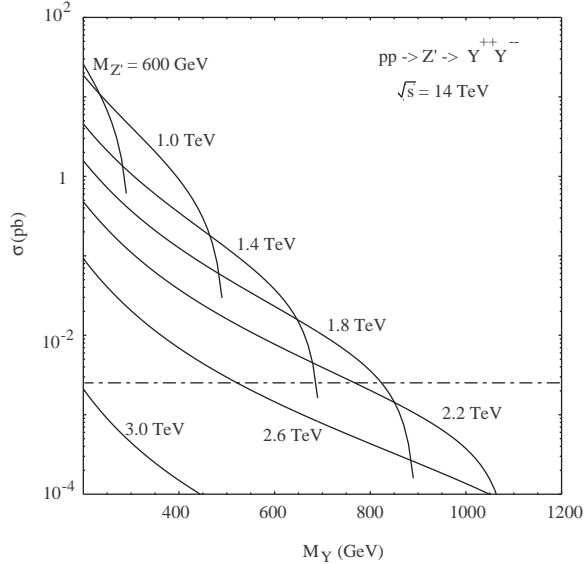,width=3.5in}}
}
\vskip-1.4truein
\caption{Real $Z'$ contribution to $Y^{++} Y^{--}$ production at the LHC
($\protect\sqrt{s} = 14$ TeV) as a function of $\MY$, for various values of
$\MZP$. The horizontal line indicates the cross section required for
discovery: 2.5 $fb$.} 
\label{Z'Y++Y--}
\end{figure}

It is instructive to separate out the contribution of an on-shell $Z'$ to
the vector bilepton production cross section from that of the $\gamma$ and
$Z$. In Fig.~\ref{Z'Y++Y--}, for various values of $\MZP$, we present the
cross section for $pp \to Y^{++} Y^{--}$ at the LHC due to the real $Z'$
alone. By comparing Figs.~\ref{YY} and \ref{Z'Y++Y--}, we can see for which
values of $\MZP$ the on-shell $Z'$ dominates the process, and for which
values it is negligible.

In particular, we note that real $Z'$ exchange is dominant only for $\MZP
\lsim 1.0$ TeV. Thus, for such values of $\MZP$, the production of
bileptons of mass $\MY \lsim 500$ GeV is due principally to the exchange of
an on-shell $Z'$. On the other hand, the real $Z'$ contribution is
basically negligible for $\MZP \gsim 1.8$ TeV, so that bileptons of mass
$\MY \gsim 900$ GeV are produced mainly via $\gamma$ or $Z$ exchange. For
1.0 TeV $\lsim \MZP \lsim$ 1.8 TeV, the $Z'$ and $\gamma,Z$ contributions
are similar in size.

We can therefore conclude that, for the entire range of $\MY$ for which
bileptons are observable at the LHC, namely 740 GeV $\le \MY \lsim$ 1 TeV, 
the cross section is never dominated by the exchange of an on-shell $Z'$.
Indeed, the $\gamma$- and $Z$-exchange contributions dominate the cross
section for the larger values of $\MY$. Thus, although we have performed
the calculations within the 3-3-1 model, the details of this model are
largely unimportant to the conclusions. In other words, the result that
vector bileptons of mass $\MY \lsim 1$ TeV are observable at the LHC is
basically model-independent.

The production of two doubly-charged bileptons will result in an
unmistakeable signature. Each of the bileptons will decay to two same-sign
leptons, not necessarily of the same flavor, leading to an $\ell_1^+
\ell_1^+ \ell_2^- \ell_2^-$ signal, in which each pair of same-sign leptons
has the same invariant mass. The SM background to this process is tiny.
Should bileptons be produced at the LHC, there should be no difficulty in
detecting them.

\subsection{$u{\bar d} \to Y^{++} Y^{-}$, ${\bar u}d \to Y^+Y^{--}$}

In the previous subsection, we noted that the process $q{\bar q} \to Y^{++}
Y^{--}$, in which the bileptons decay to $\ell_1^+ \ell_1^+ \ell_2^-
\ell_2^-$, has virtually no SM background. In fact, even if a single
doubly-charged bilepton were produced in a reaction, there would be little
background, since no SM process will give two same-sign leptons whose
invariant mass has a peak at the bilepton mass. Thus, it is also of
interest to examine processes in which one doubly-charged bilepton is
produced. We therefore consider the reactions $u{\bar d} \to Y^{++} Y^{-}$
and ${\bar u}d \to Y^+Y^{--}$ \cite{Nandi}.

The process $u{\bar d} \to Y^{++} Y^{-}$ is quite similar to $u{\bar u} \to
Y^{++} Y^{--}$. Indeed, the amplitude-squared for $u{\bar d} \to Y^{++}
Y^-$ is given by the expressions in Eqs.~\ref{uuamp1}-\ref{uuamp3}, with
the following changes: (i) there is only 1 $s$-channel diagram, with an
internal $W$, instead of 3 $s$-channel diagrams ($N,N' = \gamma,Z,Z'$),
(ii) $c_{\sss N}^{u_\lft} \to c_{\sss W} = {1\over\sqrt{2}}$, (iii)
$c_{\sss N}^{u_\rht} \to 0$, (iv) $c_{\sss N}^{\sss Y} \to c_{\sss W}^{\sss
Y} = {1\over\sqrt{2}}$. The amplitude for ${\bar u}d \to Y^+Y^{--}$ is
identical.

The cross sections for $Y^{++} Y^-$ and $Y^{--} Y^+$ production at
hadron colliders are shown in Fig.~\ref{YY}, in which we assume that
$M_{\sss Y^+} = M_{\sss Y^{++}} \equiv \MY$. At the Tevatron, which is
a $p{\bar p}$ collider, these two cross sections are equal. However,
even with the increased luminosity (and slight increase in energy) of
run 2, these processes are unobservable. At the LHC, the cross
sections for these two final states are not equal since the LHC is a
$pp$ collider, and hence has more $u$-quarks than $d$-quarks, thus
favoring the $Y^{++} Y^-$ final state. The processes $u{\bar d} \to
Y^{++} Y^{-}$ and ${\bar u}d \to Y^+Y^{--}$ are observable for $\MY
\lsim 900$ GeV and $\MY \lsim 660$ GeV, respectively. Given the
low-energy constraint of $\MY > 740$ GeV, this implies that there is a
small window of observability at the LHC for $Y^{++} Y^-$ production.

\subsection{$q{\bar q} \to Yee$}

The final process involving vector bileptons that we consider is the
reaction $q{\bar q} \to Y^{++} e^-e^-$, in which the $e^- e^-$ pair comes
from a virtual bilepton. The advantage of this process over $q{\bar q} \to
Y^{++} Y^{--}$ is clear: it is energetically easier to produce one real
bilepton than two. However, there is also a hefty price to pay -- the
amplitude involves an additional gauge coupling, and one has to consider
3-body final-state phase space instead of 2-body phase space. The only
conceivable way to offset this is if the process is dominated by the decay
of a real $Z'$, with $\MY > \MZP/2$. (Of course, if $\MY < \MZP/2$, then
pair production of vector bileptons will dominate.) The $Z'$ contribution
to the cross section for this process involves a factor
\[
{1 \over ({\hat s}-\MZP^2)^2 + (\MZP\GamZP)^2} ~.
\]
In this case, for ${\hat s} = \MZP^2$, it is perhaps possible that the
enhancement due to the on-shell $Z'$ might compensate for the above
suppressions. This is what we investigate here.

\begin{figure}[t]
\centerline{
\hskip-1truecm
\mbox{\psfig{figure=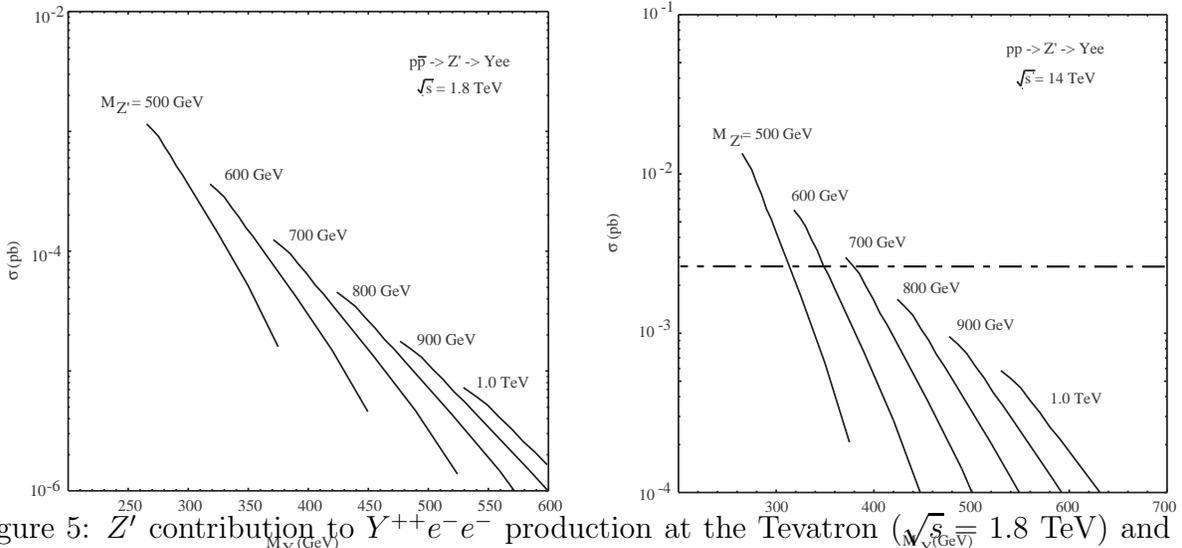,width=7.0in}}
}
\vskip-5.6truein
\caption{$Z'$ contribution to $Y^{++} e^- e^-$ production at the Tevatron
($\protect\sqrt{s} = 1.8$ TeV) and at the LHC ($\protect\sqrt{s} = 14$ TeV)
as a function of $\MY$, for various values of $\MZP$. The horizontal line
indicates the cross section required for discovery at the LHC: 2.5 $fb$.
The process is unobservable at the Tevatron.} 
\label{Yee}
\end{figure}

We therefore calculate the cross section for $q{\bar q} \to Y^{++} e^-e^-$,
mediated solely by an on-shell $Z'$. The results are shown in
Fig.~\ref{Yee}. It is clear that the reaction $q{\bar q} \to Y^{++} e^-e^-$
is completely unobservable at the Tevatron. At the LHC, depending on the
value of $\MZP$, this process is observable for $\MY \lsim 380$ GeV.
However, this range of bilepton masses has already been ruled out. And even
if such masses were still allowed, the cross section for $q{\bar q} \to
Y^{++} Y^{--}$ due only to intermediate $\gamma$ and $Z$ exchange is still
roughly two orders of magnitude larger. Thus, this process cannot be used
to discover the bileptons of the 3-3-1 model.

The problem here is that the $Z'$ in the 3-3-1 model is a relatively broad
resonance. For example, the width of a 1 TeV $Z'$ is about 200 GeV. Thus,
the hoped-for enhancement due to an on-shell $Z'$ is fairly minimal.
However, in a model in which the $Z'$ is quite narrow, say $\GamZP \sim
10^{-2} \MZP$, then the enhancement factor could be substantial, and could
well overcome the suppressions mentioned above. Indeed, in such a model,
bileptons would be more easily discovered via $q{\bar q} \to Z' \to Y^{++}
e^-e^-$ than via $q{\bar q} \to \gamma,Z \to Y^{++} Y^{--}$. Thus, we
conclude that, although the process $q{\bar q} \to Y^{++} e^-e^-$ is of
little interest within the 3-3-1 model, it might be important in other
models containing bileptons. It is therefore worthwhile to search for
signals of such a process.


\section{Scalar Bileptons}

In this section we consider the production of scalar bileptons $\phi$ at
hadron colliders. In contrast to vector bileptons, if one computes the
cross section for $q{\bar q} \to \phi^{++} \phi^{--}$ including only the
$s$-channel contributions from the $\gamma$ and $Z$, one finds that
unitarity is not violated. Thus, it is not necessary to perform the
calculations for scalar bilepton production within a particular model.

However, in a given model, there may be new, exotic contributions to
processes such as $q{\bar q} \to \phi^{++} \phi^{--}$, such as the exchange
of a $Z'$. In fact, as we will see, if one includes these additional
contributions, the cross section for scalar bilepton production may be
significantly increased relative to the case where only $\gamma$ and $Z$
exchange are considered.

Thus, if one wants to study scalar bilepton production at hadron colliders,
it is useful to examine both scenarios. In the following subsections we
therefore consider two situations concerning the process $q{\bar q} \to
\phi^{++} \phi^{--}$: (i) only the SM $\gamma$ and $Z$ contributions are
present, and (ii) there are additional, exotic contributions. For this
latter possibility, we must choose a particular model in which to perform
the calculation. As before, in this case we opt for the 3-3-1 model.

We also consider the processes $u{\bar d} \to \phi^{++} \phi^{-}$ and
${\bar u} d \to \phi^{--} \phi^{+}$, in which a single doubly-charged
scalar bilepton is produced.

Similar to the case of vector bileptons, the data on muonium-antimuonium
conversion constrain doubly-charged scalar bileptons to satisfy $M_\phi >
2$--$3.3 \, \lambda$ TeV \cite{Cuypers}. However, there is an important
difference between vector and scalar bileptons. For vector bileptons, the
coupling $\lambda$ is a gauge coupling, and is specified within a
particular model. But for scalar bileptons, $\lambda$ is the (unspecified)
Yukawa coupling of the bilepton to leptons. In the processes considered
below, the coupling $\lambda$ does not appear, and hence can be taken to be
as small as desired. Thus, although we take $M_\phi > 200$ GeV, the only
real ($\lambda$-independent) constraint on the mass of the scalar bilepton
comes from experiments at LEP, namely that it must be greater than $M_{\sss
Z}/2$.

\subsection{$q{\bar q} \to \phi^{++} \phi^{--}$}

The process $q{\bar q} \to \phi^{++} \phi^{--}$ is mediated principally by
the exchange of a neutral gauge boson $N$ ($N = \gamma, Z$ and possibly
$Z'$). (In a particular model, there may also be contributions from exotic
quarks in $t$-channel, which depend on the Yukawa coupling $\lambda$. But
since we are assuming that this coupling is small, we can ignore these
contributions.) The amplitude-squared for this process is 
\beq
\label{scalarprod}
{1\over 4}\sum_{\rm spins} |{\cal M}_1|^2 = \sum_{\sss N,N'} {1\over 4}
\, {g^4 \, c_{\sss N}^\phi \, c_{\sss N'}^\phi \, ( c_{\sss
N}^{q_\lft} \, 
c_{\sss N'}^{q_\lft} + c_{\sss N}^{q_\rht} \, c_{\sss N'}^{q_\rht} ) \over
({\hat s} - M_{\sss N}^2 + i \Gamma_{\sss N} M_{\sss N})
({\hat s} - M_{\sss N'}^2 - i \Gamma_{\sss N'} M_{\sss N'}) } \, 
{\hat s}^2 \beta^2 \sin^2\theta ~,
\eeq
where
\bea
c_\gamma^\phi & = & Q \sin\theta_w ~, \nn \\
c_{\sss Z}^\phi & = & {1\over \cos\theta_w} \, (I_3 - Q
\sin^2\theta_w) ~.
\eea
In the above, the scalar dilepton charge $Q$ can be $+2$ or $-2$, and its
weak isospin $I_3$ can in principle take any integer or half-integer
value. In the 3-3-1 model, there is also a contribution from an $s$-channel
$Z'$: 
\beq
c_{\sss Z'}^\phi = {1\over \cos\theta_w} \left[ - {
\sqrt{1-4\sin^2\theta_w} \over 2 \sqrt{3} } \, Y + { 1 - \sin^2\theta_w
\over \sqrt{3} \sqrt{1-4\sin^2\theta_w} } \, X \right],
\eeq
where $Y = 2(Q-I_3)$ is the ordinary SM hypercharge, and $X$ is the
$U(1)_{\sss X}$ charge.

Since the scalar bileptons are defined only by their quantum numbers $Q$,
$I_3$ and possibly $X$, there are an infinite number of possible cases one
can consider. The minimal 3-3-1 model contains 2 bileptons: (i)
$\eta_1^{++}$, which has $Q=+2$, $I_3 = +1$, $Y=+2$ and $X=0$, (ii)
$\eta_2^{--}$, which has $Q=-2$, $I_3 = 0$, $Y=-4$ and $X=0$. For
simplicity, in this paper we focus on the $\eta_1$ (the results for the
$\eta_2$ are quite similar).

\begin{figure}[t]
\centerline{
\hskip-1truecm
\mbox{\psfig{figure=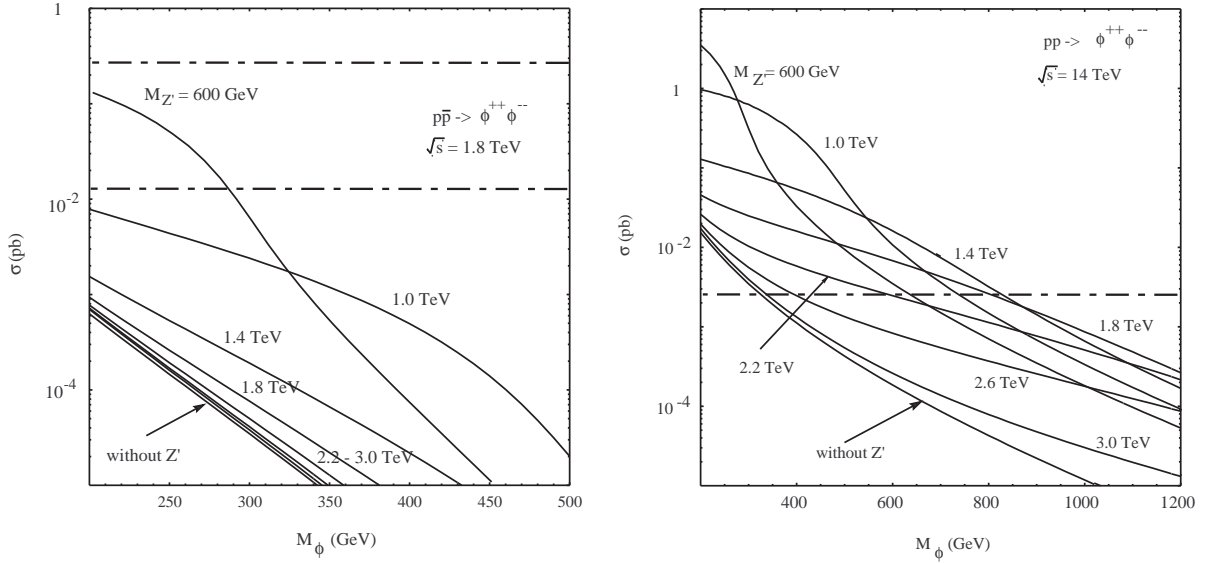,width=7.0in}}
}
\vskip-4.8truein
\caption{Cross sections for $\phi^{++} \phi^{--}$ production
($\phi=\eta_1$) at the Tevatron ($\protect\sqrt{s} = 1.8$ TeV) and at the
LHC ($\protect\sqrt{s} = 14$ TeV) as a function of $M_\phi$, for various
values of $\MZP$. The horizontal lines indicate the cross sections required
for discovery: 0.25 $pb$ (Tevatron, run 1), 12.5 $fb$ (Tevatron, run 2),
and 2.5 $fb$ (LHC).}  
\label{eta1++eta1--}
\end{figure}

In Fig.~\ref{eta1++eta1--} we show the cross sections for $\eta_1^{++}
\eta_1^{--}$ production at the Tevatron and LHC. We consider the case where
only the SM $\gamma$ and $Z$ contribute, as well as the case where there is
an additional $Z'$ contribution. It is clear from this figure that the
effect of the $Z'$ can be considerable. Depending on its mass, the cross
section can be increased by up to two orders of magnitude. (Similar
behavior was found in Ref.~\cite{bilepLRS}, where the production at
high-energy colliders of the scalar bilepton of the left-right symmetric
model was studied.)

Even with such an enhancement, this process is observable at the
Tevatron only if the particles are light: $\MZP \lsim 600$ GeV and
$M_\phi \lsim 300$ GeV are required to obtain an observable signal.
(This holds for both $\sqrt{s} = 1.8$ and 2.0 TeV.) Since the $Z'$ is
already constrained to satisfy $\MZP > 550$ GeV, this does not leave
much room. On the other hand, $\eta_1^{++} \eta_1^{--}$ production is
observable at the LHC for larger masses. But the reach in $M_\phi$
depends strongly on whether a $Z'$ is present, and if so, what the
value of its mass is. If there is no $Z'$, then such scalar bileptons
can be observed for $M_\phi \lsim 375$ GeV. If a $Z'$ is present, then
this reach increases to 800-850 GeV.

Of course, these results apply specifically to the $\eta_1^{++}$ bilepton.
For a different given bilepton, the observational reach will depend on its
quantum numbers. In particular, for bileptons other than the $\eta_1^{++}$
it is conceivable that this reach can extend to higher masses. Still, the
results we have found for the $\eta_1^{++}$ give one a feel for the order
of magnitude of the reach that one can expect for generic scalar bileptons.

We therefore conclude that the process $q{\bar q} \to \phi^{++} \phi^{--}$
is observable at the Tevatron only if the $\phi^{++}$ is light and if the
mass of the $Z'$ lies just above its present bound. For the LHC, if only
the $\gamma$ and $Z$ contribute to this process, then bileptons are
observable if $M_\phi \lsim 400$ GeV. And if the contributions from non-SM
particles such as $Z'$ bosons are significant, then this limit may be
pushed up to $M_\phi \lsim 1$ TeV.

We must also reiterate that these results are independent of $\lambda$, the
Yukawa coupling of the scalar bilepton to leptons. This is in contrast to
Bhabha and M{\o}ller scattering at $e^+e^-$ and $e^-e^-$ colliders.
Although these lepton colliders are potentially sensitive to much larger
scalar bilepton masses, their reach depends directly on the value of the
Yukawa coupling. If this coupling is too small, then there will be no
measurable effect in $e^+e^-$ and $e^-e^-$ colliders, but the process
$q{\bar q} \to \phi^{++} \phi^{--}$ will still be observable. (Of course,
the process $e^+ e^- \to \phi^{++} \phi^{--}$, which is also independent of
$\lambda$, may be possible, depending on $M_\phi$ and $\sqrt{s}$.) Thus, if
the Yukawa coupling is small, a hadron collider such as the LHC may in fact
be the optimal machine for detecting scalar bileptons.

\subsection{$u{\bar d} \to \phi^{++} \phi^{-}$, ${\bar u} d \to \phi^{+}
\phi^{--}$}

We also consider the production of a single doubly-charged scalar dilepton
via $u{\bar d} \to \phi^{++} \phi^{-}$ or ${\bar u} d \to \phi^{+}
\phi^{--}$. These processes are mediated by the exchange of a $W$ in
$s$-channel. As in the vector case, this process is similar to $q {\bar q}
\to \phi^{++} \phi^{--}$, assuming that the masses of the singly-charged
and doubly-charged dileptons are equal. (Due to constraints from the
$\rho$-parameter, these masses cannot be too different.) The
amplitude-squared of these processes is the same as Eq.~\ref{scalarprod},
with the following changes: (i) $q{\bar q}$ becomes $u{\bar d}$, (ii)
$c_{\sss N}^{q_\lft} \to {1\over \sqrt{2}}$, (iii) $c_{\sss N}^{q_\rht} \to
0$, (iv) $c_{\sss N}^\phi \to c_{\sss W}^\phi$. Therefore, for the process
$u{\bar d} \to \phi^{++} \phi^-$ or ${\bar u} d \to \phi^{+} \phi^{--}$, we
have
\beq
{1\over 4}\sum_{\rm spins} |{\cal M}|^2 = {1\over 8}
\, {g^4 \, (c_{\sss W}^\phi)^2 \over
({\hat s} - M_{\sss W}^2)^2 + (\Gamma_{\sss W} M_{\sss W})^2 }
\, {\hat s}^2 \beta^2 \sin^2\theta ~.
\eeq

The quantity $c_{\sss W}^\phi$ parametrizes the
$W^-$--$\phi^{++}$--$\phi^-$ coupling. Its value depends on the
representation that the scalar dileptons are in, as well as how they are
defined. For example, for an ordinary $SU(2)_\lft$-doublet, $c_{\sss
W}^\phi = {1\over\sqrt{2}}$. In the minimal 3-3-1 model, only the
$\eta_1^{++}$ couples to the $W$, with $c_{\sss W}^\phi = 1$.

\begin{figure}[t]
\centerline{
\hskip-1truecm
\mbox{\psfig{figure=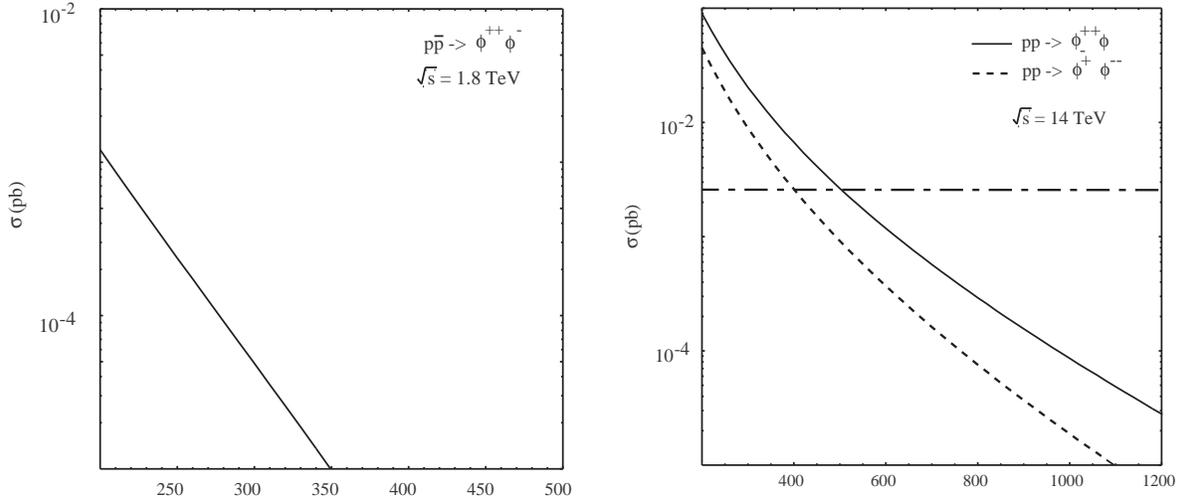,width=7.0in}}
}
\vskip-5.6truein
\caption{Cross sections for $\phi^{++} \phi^{-}$ and $\phi^{+} \phi^{--}$ 
production ($\phi^{++}=\eta_1^{++}$) at the Tevatron ($\protect\sqrt{s} =
1.8$ TeV) and at the LHC ($\protect\sqrt{s} = 14$ TeV) as a function of
$M_\phi$. The horizontal line indicates the cross section required for
discovery at the LHC: 2.5 $fb$. The process is unobservable at the
Tevatron.}
\label{phi++phi-}
\end{figure}

In Fig.~\ref{phi++phi-} we present the cross sections for $\phi^{++}\phi^-$
and $\phi^{+}\phi^{--}$ production at the Tevatron and at the LHC.
Specifically, we consider $\phi^{++} = \eta_1^{++}$. (The case of an
$SU(2)_\lft$-doublet, or indeed any other $SU(2)_\lft$ representation, can
be obtained by simply scaling the results of Fig.~\ref{phi++phi-} by
$(c_{\sss W}^\phi)^2$.) This process is unobservable at the Tevatron, but
can be observed at the LHC for scalar bilepton masses $M_\phi \lsim 400$
GeV ($\phi^{+}\phi^{--}$) or $M_\phi \lsim 500$ GeV ($\phi^{++}\phi^{-}$).

It is also conceivable that, in a particular model, there might be new,
exotic contributions to $q{\bar q} \to \phi^{++} \phi^{-}$. If these
include the exchange of an on-shell particle in $s$-channel, then the cross
section may be significantly enhanced relative to the results of
Fig.~\ref{phi++phi-}. As we saw in the case of $q{\bar q} \to \phi^{++}
\phi^{--}$, this enhancement may be large, perhaps as much as several
orders of magnitude. Although this does not happen in the 3-3-1 model for
$q{\bar q} \to \phi^{++} \phi^{-}$, it may occur in other models. So this
possibility should be kept in mind.


\section{Conclusions}

We have investigated the production of vector and scalar bileptons at
hadron colliders. Although we would have liked to perform this analysis in
a model-independent way, this is unfortunately not possible. To see this, 
consider, for example, the process $q{\bar q} \to Y^{++} Y^{--}$, where
$Y^{++}$ is a vector bilepton. If one considers only the SM contributions
($\gamma$ and $Z$ exchange) to this process, one finds that the cross
section violates unitarity. In order to avoid unitarity violation, it is
therefore necessary that the calculations for vector bilepton production be
performed within a specific model.

We have chosen the 3-3-1 model. In this model unitarity is restored by the
inclusion of additional contributions to $q{\bar q} \to Y^{++} Y^{--}$:
$s$-channel $Z'$ exchange and $t$-channel exotic $D$-quark exchange.
Although, strictly speaking, all our results are model-dependent, we do
expect the cross sections to take similar (order-of-magnitude) values in
most models. Indeed, for certain values of the $Z'$ and $Y$ masses, we find
that some processes are in fact dominated by the SM contributions. The
results for these processes are thus essentially model-independent.

Processes involving scalar bileptons $\phi^{++}$, such as $q{\bar q} \to
\phi^{++} \phi^{--}$, do not suffer from unitarity violation, so it is not
necessary to use a particular model. However, in a given model, there may
be new, exotic contributions to scalar bilepton production which may
significantly increase the cross sections. It is therefore useful to
calculate the cross sections for scalar bilepton production within a chosen
model, as well as using only the SM ($\gamma$, $Z$) contributions. For the
model-dependent calculations, we have again used the 3-3-1 model, focussing
on one of its bileptons, the $\eta_1^{++}$.

There are constraints on bileptons from low-energy experiments. The data on
muonium-antimuonium conversion imply a lower limit on the mass of the
doubly-charged vector bileptons of the 3-3-1 model: $\MY > 740$ GeV. For
singly-charged vector bileptons, constraints from $\mu_\rht \to e\nu\nu$
yield a somewhat weaker limit: $\MY > 440$ GeV. For scalar bileptons, the
data on muonium-antimuonium conversion require $M_\phi > 2$--$3.3 \,
\lambda$ TeV, where $\lambda$ is the Yukawa coupling of the bilepton to
leptons. But $\lambda$ is unknown, and can be taken as small as desired.
Thus, the only real constraint on the scalar bilepton mass is $M_\phi >
M_{\sss Z}/2$.

We have examined a variety of processes in which one or two
doubly-charged bileptons are produced at a hadron collider. We
considered both the Tevatron ($\sqrt{s}=1.8$ TeV) and the LHC
($\sqrt{s}=14$ TeV). In all cases, as a figure of merit, we required
25 events for observability. This corresponds to a cross section of
0.25 $pb$ (Tevatron, run 1), 12.5 $fb$ (Tevatron, run 2), or 2.5 $fb$
(LHC). (Even with $\sqrt{s} = 2.0$ TeV for run 2 at the Tevatron, our
results are little changed.)

Note that if a doubly-charged vector or scalar bilepton were produced at a
collider, its decay would yield an unmistakeable signature. The decay
products would be a pair of same-sign leptons, with an invariant mass equal
to that of the bilepton. The SM background to such a process is very small.
Should bileptons be produced at a hadron collider, there should be no
difficulty in detecting them.

Here are our results:

\begin{itemize}

\item $q{\bar q} \to Y^{++} Y^{--}$: At the Tevatron, vector bileptons
are observable if $\MY \lsim 250$ GeV (run 1) or $\MY \lsim 320$ GeV 
(run 2) [$\MY \lsim 360$ GeV (run 2, $\sqrt{s} = 2.0$ TeV)]. 
However, in all cases this mass range has already been ruled out, so 
we conclude that vector bileptons cannot be observed in this process 
at experiments at the Tevatron. At the LHC, bileptons of mass $\MY 
\lsim 1$ TeV are observable. In this case there is a window of 
observability. Note also that bileptons of mass $\MY \gsim 900$ GeV 
are produced mainly via $\gamma$ or $Z$ exchange -- the $Z'$ of the 
3-3-1 contributes little. Thus, the LHC result is largely 
model-independent.

\item $u{\bar d} \to Y^{++} Y^{-}$, ${\bar u}d \to Y^+Y^{--}$: Given the
low-energy constraint of $\MY > 740$ GeV, neither of these processes is
observable at the Tevatron. At the LHC, only the process $u{\bar d} \to
Y^{++} Y^{-}$ is observable, for $\MY \lsim 900$ GeV.

\item $q{\bar q} \to Yee$: Suppose that the process $q{\bar q} \to Y^{++}
Y^{--}$ were dominated by the exchange of a real $Z'$ (this obviously
requires $\MY < \MZP/2$). Consider now the case where $\MY > \MZP/2$. Here,
it is conceivable that the process $q{\bar q} \to Y^{++} e^-e^-$ is
observable while $q{\bar q} \to Y^{++} Y^{--}$ is not. We have therefore
calculated the cross section for $q{\bar q} \to Y^{++} e^-e^-$, mediated
solely by an on-shell $Z'$. Unfortunately, given the present low-energy
constraints $\MY$, we find that this process is unobservable at both the
Tevatron and the LHC.

However, this result is highly model-dependent. In the 3-3-1 model, the
$Z'$ is a broad resonance, so that its on-shell production yields little
enhancement of the cross section. But if the width of the $Z'$ were narrow,
say $\GamZP \sim 10^{-2} \MZP$, as could be the case in another model, then
the enhancement factor could be substantial. In such a case, bileptons
would be more easily discovered via $q{\bar q} \to Y^{++} e^-e^-$ than via
$q{\bar q} \to Y^{++} Y^{--}$. It is therefore worthwhile to search for
signals of $q{\bar q} \to Yee$.

\item $q{\bar q} \to \phi^{++} \phi^{--}$: We have calculated this cross
section for $\phi^{++} = \eta_1^{++}$ (the $\eta_1^{++}$ has quantum
numbers $Q=+2$, $I_3 = +1$, $Y=+2$). We considered (i) the case where only
the SM $\gamma$ and $Z$ contribute, as well as (ii) the case where there is
an additional $Z'$ contribution. The effect of the $Z'$ can be
considerable: depending on its mass, the cross section can be increased by
up to two orders of magnitude.

Even so, this process is barely observable at the Tevatron -- an observable
signal requires fairly light particles: $\MZP \lsim 600$ GeV and $M_\phi
\lsim 300$ GeV (and note that $\MZP$ is already constrained to be above 550
GeV). The LHC is considerably more promising, but the reach in $M_\phi$
depends strongly on whether a $Z'$ is present. If there is no $Z'$, then
such scalar $\eta_1^{++}$ bileptons can be observed for $M_\phi \lsim 375$
GeV, while if a $Z'$ is present, then this reach increases to 800-850 GeV.

Note that, although these results have been calculated specifically for the
$\eta_1^{++}$ bilepton, they should also apply, to within factors of order
1, to all scalar bileptons, as long as the bilepton's quantum numbers are
not too unconventional. Furthermore, we reiterate that, in contrast to 
Bhabha and M{\o}ller scattering at $e^+e^-$ and $e^-e^-$ colliders, our
results are independent of $\lambda$, the Yukawa coupling of the scalar
bilepton to leptons.

\item $u{\bar d} \to \phi^{++} \phi^{-}$, ${\bar u} d \to \phi^{+}
\phi^{--}$: We take $\phi^{++}$ to have $I_3 = +1$. These processes
are again unobservable at the Tevatron, but can be observed at the LHC for
scalar bilepton masses $M_\phi \lsim 400$ GeV ($\phi^{+}\phi^{--}$) or
$M_\phi \lsim 500$ GeV ($\phi^{++}\phi^{-}$). Note also that this reach can
be considerably increased in models in which there is an additional
contribution due to the $s$-channel exchange of an on-shell particle.

\end{itemize}

To summarize, vector bileptons are unobservable at experiments at the
Tevatron, while there is only a small window for detection of scalar
bileptons. On the other hand, depending on the process, vector and scalar
bileptons of masses up to about 1 TeV may be observable at the LHC.

\bigskip
\centerline{\bf Acknowledgements}
\bigskip

This research was financially supported by NSERC of Canada and FCAR du
Qu\'ebec.

\end{document}